\documentclass[%
 aip,
 jmp,%
 amsmath,amssymb,
 reprint,%
]{revtex4-1}

\usepackage{graphicx}
\usepackage{dcolumn}
\usepackage{bm}
\begin{document}

\title[Cosmology with Ricci-type dark energy]{Cosmology with Ricci-type dark energy}

\author{W. Zimdahl}

\affiliation{Departamento de
F\'{\i}sica, Universidade Federal do Esp\'{\i}rito Santo\\
Vit\'{o}ria, Esp\'{\i}rito Santo, Brazil}

\author{J.C. Fabris}%
\affiliation{Departamento de
F\'{\i}sica, Universidade Federal do Esp\'{\i}rito Santo\\
Vit\'{o}ria, Esp\'{\i}rito Santo, Brazil}%

\author{S. del Campo}
\affiliation{Instituto de F\'{\i}sica,
Pontificia Universidad Cat\'{o}lica de Valpara\'{\i}so, Chile}%

\author{R. Herrera}
\affiliation{Instituto de F\'{\i}sica,
Pontificia Universidad Cat\'{o}lica de Valpara\'{\i}so, Chile}%
\date{\today}
\begin{abstract}
We consider the dynamics of a cosmological substratum of pressureless matter and holographic dark energy with a cutoff length proportional to the Ricci scale.
Stability requirements for the matter perturbations are shown to single out a model with a fixed relation between the present matter fraction $\Omega_{m0}$ and the present value $\omega_{0}$ of the equation-of-state parameter of the dark energy. This model has the same number of free parameters as the $\Lambda$CDM model but it has
no $\Lambda$CDM limit.  We discuss the consistency between background observations and the mentioned
stability-guaranteeing parameter combination.
\end{abstract}



\maketitle

\section{\label{sec:introduction}Introduction}

Among the alternative approaches to describe the dark cosmological sector, consisting of dark matter (DM) and dark energy (DE), so called holographic DE  models have received considerable attention \cite{cohen,li}.
The underlying holographic principle states that the number of
degrees of freedom in a bounded system should be finite and related to the area of its boundary \cite{tH}.
On this basis, a field theoretical relation between a short
distance (ultraviolet) cutoff and a long distance (infrared)
cutoff was established \cite{cohen}. Such relation ensures that  the energy in a box of size $L$ does not
exceed the energy of a black hole of the same size.
If applied to the dynamics of the Universe, $L$ has to be a
cosmological length scale. Different choices of this cutoff scale result in different DE
 models.
For the most obvious choice, the Hubble scale, only models in which DM and DE are interacting with each other also nongravitationally, give rise to a suitable dynamics \cite{DW,HDE}. Following Ref.~\onlinecite{li},
there has been a considerable number of investigations based on the future event horizon as cutoff scale.  All models with a cutoff at the future event horizon, however, suffer from the serious drawback that they cannot describe a transition from decelerated to accelerated expansion. A future event horizon does not exist during the period of decelerated expansion.
We are focussing here on a further option that has received attention more recently, a model based on a cutoff length proportional to the Ricci scale.
A distance proportional to the Ricci scale has been identified as a causal connection scale for
perturbations \cite{brustein}. As a cutoff length in DE
models it was first used in Ref.~\onlinecite{gao}.
Subsequent investigations include Ref.~\onlinecite{cai} and, for the perturbation dynamics,
Refs.~\onlinecite{fengli,karwan,yuting}.\\
Here, we reconsider the dynamics of a two-component system of pressureless DM and Ricci-type DE
both in the homogeneous and isotropic background and on the perturbative level\cite{SRJW,SRJW2}.
Whereas most dynamic DE scenarios start with an assumption for the equation-of-state (EoS) parameter for the DE, the starting point of holographic models is an expression for the DE energy density from which the EoS is then derived. As was pointed out in Ref.~\onlinecite{SRJW}, the mere definition
 of the holographic DE density, independently of the choice of the specific cutoff length, implies an interaction with the DM component.
 Requiring this interaction to vanish is equivalent to impose an additional condition on the dynamics.
 In the case of Ricci-type DE this condition establishes a simple relation between the matter fraction and the necessarily time-dependent EoS parameter. Of course, a time-varying EoS parameter is not compatible with a cosmological constant.
Our main aim here is to perform a gauge-invariant perturbation analysis for this model.
It will turn out that the general perturbation dynamics suffers from instabilities. There exists just one single configuration without instabilities at finite values of the scale factor $a$ \cite{karwan,SRJW2}.
We also update previous tests of the homogeneous and isotropic background dynamics using recent results for
the differential age of old objects based on the $H(z)$ dependence, data from SNIa  and from BAO.

\section{\label{sec:hde}Holographic Dark Energy}

\subsection{General properties}

The cosmic medium is assumed to consist of pressureless DM with energy density
$\rho_{m}$ and a holographic DE component with energy density $\rho_{H}$.
In the spatially flat case Friedmann's equation is
\begin{equation}
3 H^2 = 8\pi\, G (\rho_{m} + \rho_{H})  \,,\label{Fried1}
\end{equation}
where $H=\frac{\dot{a}}{a}$ is the Hubble rate and $a$ is the scale factor of the Robertson-Walker metric.
In general, both components are not separately conserved  but obey the balance equations
\begin{equation}
\dot{\rho}_{m} + 3 H \rho_{m}  = Q\,,\quad
\dot{\rho}_{H} + 3 H (1 + \omega)\rho_{H} = -Q \, \label{cons2}
\end{equation}
with a source (or loss) term $Q$,
such that the total energy $\rho = \rho_{m} + \rho_{H}$ is conserved.
Here,
$\omega \equiv\frac{p_H}{\rho_{H}} = \frac{p}{\rho_{H}}$ is the EoS parameter of the DE and
$p_H$ is the pressure associated with the holographic component.
In terms of the scale factor as independent variable, the acceleration equation can be written
\begin{equation}
\frac{d\ln H}{d\ln a}  = - \frac{3}{2}\left(1 + \frac{\omega(a)}{1 + r(a)}\right)\,,
\label{dH}
\end{equation}
where $r\equiv\frac{\rho_{m}}{\rho_{H}}$ is the ratio of the energy densities.
The \textit{total} effective EoS of the cosmic medium is
\begin{equation}
\frac{p}{\rho} = \frac{\omega}{1 + r}  \,. \label{w}
\end{equation}
According to the balance equations (\ref{cons2}), the ratio $r$ changes as
\begin{equation}
\dot{r} = 3Hr\,\left(1 + r\right)\,\left[\frac{\omega}{1+r} + \frac{Q}{3H\rho_{m}}\right]\ . \label{dr2}
\end{equation}
Following Refs.~\onlinecite{cohen,li}, we write the holographic energy density as
\begin{equation}
\rho_H=\frac{3\,c^2\,M_p^2}{L^{2}} \,.\label{ans}
\end{equation}
The quantity $L$ is the infrared cutoff scale and
$M_p=1/\sqrt{8\pi\,G}$ is the reduced Planck mass. The numerical constant $c^{2}$ determines the degree
of saturation of the condition
\begin{equation}
L^{3}\, \rho_H\leq M_{Pl}^{2}\, L \,,    \label{DEineq}
\end{equation}
which is at the heart of any holographic DE model. It states that the energy in a box of size $L$ should not exceed the energy of a black hole of the same size \cite{cohen}.

Differentiation of the expression (\ref{ans}) for the holographic DE density and use of the energy balances (\ref{cons2}) yields
\begin{equation}
\frac{Q}{\rho_{H}} = 2\frac{\dot{L}}{L} - 3H\left(1+\omega\right) \,. \label{QL}
\end{equation}
In general, there is no reason for $Q$ to vanish. Assuming $Q=0$ provides us with a specific relationship between $\omega$ and the ratio of the rates $\frac{\dot{L}}{L}$ and $H$. Any nonvanishing $Q$ will modify this relationship.

With $Q$ from (\ref{QL}), the general dynamics (\ref{dr2}) of the energy-density ratio $r$ becomes
\begin{equation}
\dot{r} = - 3H\,\left(1 + r\right)\,\left[1 + \frac{\omega}{1+r} - \frac{2}{3} \frac{\dot{L}}{H L}\right]\,. \label{drL}
\end{equation}
The case without interaction is characterized by [cf. Eq.~(\ref{dr2})]
\begin{equation}
Q = 0 \quad \Rightarrow\quad \dot{r} = r\left(2\frac{\dot{L}}{L} - 3 H\right)
= 3 H\,r\,\omega \label{QL0}
\end{equation}
with a generally time-dependent $\omega$.

Different choices of the cutoff scale $L$ give rise to different
expressions for the total effective EoS parameter
in Eq.~(\ref{w}) and to different relations between $\omega$ and $r$.
We shall briefly sketch  the situations for the Hubble radius and for the future event horizon
as cutoff lengths before
considering in detail the Ricci scale.

\subsection{Hubble-scale cutoff}

For $L=H^{-1}$ the holographic DE density is
\begin{equation}
\rho_H= 3\,c^2\,M_p^2 \,H^{2}  \ . \label{rh}
\end{equation}
For the deceleration parameter one has
\begin{equation}
q = - 1 - \frac{\dot{H}}{H^{2}} = \frac{1}{2}\left(1 - \frac{Q}{H \rho_{m}}\right)
\ . \label{q}
\end{equation}
In the interaction-free case we recover the Einstein-de Sitter value $q=\frac{1}{2}$.
The condition for accelerated expansion is $Q > H\rho_{m}$. To describe a transition from decelerated to
accelerated expansion, $Q$ has to change from $Q < H\rho_{m}$ to $Q > H\rho_{m}$.
A viable scenario can be realized, e.g., by a choice \cite{HDE}
\begin{equation}\label{}
  \frac{Q}{3H\rho_{m}} = \mu \left(\frac{H}{H_{0}}\right)^{-n}\,,
\end{equation}
where $\mu$ is an interaction constant.
The resulting dynamics is that of a generalized Chaplygin gas with a Hubble rate
\begin{equation}
\frac{H}{H_{0}} = \left(\frac{1}{3}\right)^{1/n}\left[1 - 2q_{0} + 2\left(1 +
q_{0}\right)a^{-3n/2}\right]^{1/n}\,,
\label{Hq}
\end{equation}
where $\mu$ is related to the present value $q_{0}$ of the deceleration parameter $q$ by
\begin{equation}
\mu = \frac{1}{3}\left(1 - 2q_{0}\right)
\,.
\label{muq}
\end{equation}
For $n=2$ one reproduces the
$\Lambda$CDM dynamics.

\subsection{Event-horizon cutoff}
With $L=R_{E}$,  where
\begin{equation}
R_E(t)=a(t)\,\int_t^\infty\,\frac{dt'}{a(t')}=a\,\int_a^\infty\,\frac{da'}{H'\,a'^2}\
\label{re}
\end{equation}
is the future event horizon, the holographic DE density (\ref{ans}) is
\begin{equation}
\rho_H= \frac{3\,c^2\,M_p^2}{R_{E}^{2}}  \,.\label{rhoE}
\end{equation}
The DE balance (\ref{cons2}) can be written as
\begin{equation}
\dot{\rho}_{H} + 3 H (1 + \omega_{eff}^{E})\rho_{H} = 0
\  \label{deeff}
\end{equation}
with an effective EoS (the superscript E denotes the event horizon)
\begin{equation}
\omega_{eff}^{E} = \omega + \frac{Q}{3H\rho_{H}} = - \frac{1}{3}\left(1 + \frac{2}{R_{E}H} \right)\,.
  \label{weff}
\end{equation}
This effective EoS does not directly depend on $\omega$. However, the ratio $r$ that enters $R_{E}H$ is determined by $\omega$ via Eq.~(\ref{drL}). Notice also, that this effective EoS for the DE component is different
from the total effective EoS of the cosmic medium which is $\frac{\omega}{1+r}$. In the previous Hubble-scale-cutoff case both these quantities were identical.

Different from the previous Hubble-scale cutoff, there exists a non-interacting limit with accelerated expansion in the present case.
In this special situation
\begin{equation}\label{}
\omega = -\frac{1}{3}\,\left[1+\,\frac{2}{c\,\sqrt{1+r}}\right]\ \mathrm{and} \
R_{E}H = - \frac{2}{1 + 3 \omega}
\label{Q0}
\end{equation}
are valid.
Explicitly, the relation between $\omega$ and $r$ is
\begin{equation}
\omega = - \frac{1}{3} + \frac{1}{3}\left(1 + 3\omega_{0}\right)\sqrt{\frac{1+r_{0}}{1+r}}
\,.
\label{wrE}
\end{equation}
It is obvious, that for any $\omega_{0}\approx -1$, the parameter $\omega$ remains always smaller than $-\frac{1}{3}$, demonstrating the impossibility
of a matter-dominated period in this context.

\subsection{Ricci-scale cutoff}

Our interest in the present paper will be the Ricci-scale cutoff.
The role of a distance proportional to the Ricci scale as a causal connection scale for
perturbations was noticed in Ref.~\onlinecite{brustein}. In Ref.~\onlinecite{gao} it was used for the first time as a DE cutoff scale.
The Ricci scalar
is $R = 6\left(2H^{2} + \dot{H}\right)$.
For the corresponding cutoff scale one has $L^{2} = 6/R$, i.e.,
\begin{equation}
\rho_H= 3\,c^2\,M_p^2 \,\frac{R}{6} = \alpha\left(2H^{2} + \dot{H}\right) \ ,  \label{rhR}
\end{equation}
where $\alpha = \frac{3c^{2}}{8\pi G}$.
Upon using (\ref{dH})
we obtain
\begin{equation}
\rho_H= \frac{\alpha}{2}\,H^{2}\left(1 - 3\frac{\omega}{1+r}\right)  \label{rh2}
\end{equation}
for the holographic DE density.
Notice that the (not yet known) EoS parameter explicitly enters $\rho_H$. Use of Friedmann's equation provides us with
\begin{equation}
1 =  \frac{c^{2}}{2}\left(1 + r - 3 \omega\right) \ \Rightarrow\ \omega = \frac{1}{3}\left(1 + r\right) - \frac{2}{3c^{2}}\ ,
\label{1=}
\end{equation}
which coincides with the result in Ref.~\onlinecite{karwan}.
Obviously, a constant value of $\omega$ necessarily implies a constant $r$ and vice versa.
For the source term we have \cite{SRJW}
\begin{equation}
Q  = - \frac{3H}{1+r}\left[r \omega - \frac{\dot{\omega}}{H}\right]\rho_{H}
\,. \label{Q=}
\end{equation}
It is obvious that a constant EoS parameter $\omega$ is compatible with $Q=0$ only for
$\omega=0$, i.e., if $\rho_{H}$ behaves as dust.
For a time-varying EoS parameter $\dot{\omega}\neq 0$, however, there exists a non trivial case $Q=0$:
\begin{equation}
Q = 0 \quad \Rightarrow\quad r \omega = \frac{\dot{\omega}}{H}
\quad \Rightarrow\quad r = \frac{d\ln \omega}{d\ln a}
\,. \label{rQ0}
\end{equation}
In the remainder of the paper we shall consider this case, for which the EoS parameter is explicitly given by
\begin{equation}\label{wsol}
\omega = \omega_{0}\frac{r_{0} - 3\omega_{0}}{r_{0}a^{-(r_{0} - 3\omega_{0})}- 3 \omega_{0}}\,.
\end{equation}
At high redshift we have
\begin{equation}
\omega \rightarrow 0 \ ,\qquad r \rightarrow r_0 - 3\omega_0 \ , \qquad\qquad (a \ll 1)
\ . \label{asmall}
\end{equation}
The property that noninteracting Ricci-DE behaves as dust at high redshift was first pointed out in Ref.~\onlinecite{gao}.
This model naturally reproduces an early matter-dominated era. For $r_{0} \approx \frac{1}{3}$ and $\omega_{0} \approx -1$, the ratio $r$ approaches $r\approx \frac{10}{3}$ for $a \ll 1$. This value is only roughly ten times larger than the present value $r_{0}$.
For the $\Lambda$CDM model the corresponding difference is about nine orders of magnitude. In this sense, the coincidence problem is considerably alleviated for our Ricci CDM model for which the Hubble rate becomes
\begin{equation}
\frac{H}{H_{0}} = a^{-3/2}\,\sqrt{\frac{3\omega_{0}a^{\left(r_{0}-3\omega_{0}\right)}- r_{0}\left[1+r_{0}-3\omega_{0}\right]}
{3\omega_{0}- r_{0}\left[1+r_{0}-3\omega_{0}\right]}}
\,. \label{solHQ0}
\end{equation}

\section{\label{sec:perturbations}Perturbation Dynamics}

\subsection{The two-component system}

Generally, the two-component model is described by an energy-momentum tensor
\begin{equation}
T_{ik} = \rho u_{i}u_{k} + p h_{ik}\,, \quad T_{\ ;k}^{ik} = 0\
\label{T}
\end{equation}
with $h _{ik}=g_{ik} + u_{i}u_{k}$ and $g_{ik}u^{i}u^{k} = -1$. The quantity $u^{i}$ denotes the total four-velocity of the cosmic substratum. Latin indices run from $0$ to $3$.
The total $T_{ik}$ splits into a matter component and a holographic DE component,
\begin{equation}\label{Ttot}
T^{ik} = T_{m}^{ik} + T_{H}^{ik},
\end{equation}
with ($A= m, H$)
\begin{equation}\label{TA}
T_{A}^{ik} = \rho_{A} u_A^{i} u^{k}_{A} + p_{A} h_{A}^{ik}\,,\quad\ h_{A}^{ik} = g^{ik} + u_A^{i} u^{k}_{A} \,.
\end{equation}
For separately conserved fluids we have $T_{m\ ;k}^{ik} = 0$ and $T_{H\ ;k}^{ik} = 0$.
In general, each component has its own four-velocity, with $g_{ik}u_{A}^{i}u_{A}^{k} = -1$. For the homogeneous and isotropic background we assume $u_{m}^{a} = u_{H}^{a} = u^{a}$.

Indicating first-order perturbations about the homogeneous and isotropic background by a hat symbol, the perturbed time components of the four-velocities are
\begin{equation}
\hat{u}_{0} = \hat{u}^{0} = \hat{u}_{m}^{0} =\hat{u}_{H}^{0}  = \frac{1}{2}\hat{g}_{00}\ .
\label{u0}
\end{equation}
Restricting ourselves to scalar perturbations, we define the (three-) scalar quantities $v$, $v_{m}$ and $v_{H}$ by
\begin{equation}
\hat{u}_{\mu} \equiv v_{,\mu}\,,\quad \hat{u}_{m\mu} \equiv v_{m,\mu} \,\quad \hat{u}_{H\mu} \equiv v_{H,\mu}\,.
\label{}
\end{equation}

\subsection{Basic gauge-invariant set of equations}

Introducing fractional energy-density perturbations
$\delta \equiv \frac{\hat{\rho}}{\rho}$ and changing to gauge-invariant variables according to
\begin{equation}\label{}
\delta^{c} = \delta + \frac{\dot{\rho}}{\rho}v\,,\
\hat{\Theta}^{c} = \hat{\Theta}  + \dot{\Theta}v\,,\ \hat{p}^{c} = \hat{p}  + \dot{p}v\,,
\end{equation}
energy and momentum conservations for the cosmic medium as a whole reduce to \cite{VDF}
\begin{equation}\label{balcomb}
\dot{\delta}^{c} - \Theta \frac{p}{\rho}\delta^{c} + \left(1 + \frac{p}{\rho}\right)\hat{\Theta}^{c} = 0\ .
\end{equation}
The superscript c indicates that the corresponding variables are defined with respect to a comoving observer.
The perturbation $\hat{\Theta}^{c}$ has to be determined from the perturbed Raychaudhuri equation for $\Theta$.
In our context one finds at linear order
\begin{equation}\label{dThetacfin}
\dot{\hat{\Theta}}^{c} + \frac{2}{3}\Theta\hat{\Theta}^{c} + 4\pi G\rho\delta^{c}
 - \dot{u}^{a}_{:a} = 0\
\end{equation}
with
\begin{equation}
\dot{u}^m_{;m} =
 - \frac{1}{a^2}\frac{\Delta \hat{p}^{c}}{\rho + p}\,,
\label{}
\end{equation}
where $\Delta$ is the three-dimensional Laplacian.
Combing Eqs.~(\ref{balcomb}) and (\ref{dThetacfin}) and specifying the pressure perturbations
will result in a second-order equation for $\delta^{c}$.
On the other hand,
with $\delta_{H} \equiv \frac{\hat{\rho}_{H}}{\rho_{H}}$ and
\begin{equation}\label{}
D_{H} \equiv \frac{\hat{\rho}_{H}}{\rho_{H} + p_{H}} = \frac{\delta_{H}}{1 + \omega}\
\end{equation}
it will be useful to consider the combination $S_{mH} \equiv \delta_{m} - D_{H}$, where $\delta_{m} \equiv \frac{\hat{\rho}_{m}}{\rho_{m}}$.
Our aim is to obtain an equation also for $S_{mH}$. To this purpose we have to introduce a further ingredient.
So far the pressure perturbations are not sufficiently specified.
In general, pressure perturbations in two-component systems are nonadiabatic.
Firstly, because of the two-component nature itself, secondly because each of the components may be nonadiabatic
on its own. The relevant combination for our DE component is
\begin{equation}\label{hatpH-1}
\hat{p}_{H}-
\frac{\dot{p}_H}{\dot{\rho}_H}\hat{\rho}_{H}  = \rho_{H}\left[\hat{\omega} + \frac{\omega}{3} r D_{H}\right]\,,
\end{equation}
which is a gauge-invariant expression.
Now an assumption for the perturbed EoS parameter $\hat{\omega}$ is necessary to proceed. We shall restrict ourselves here to adiabatic internal perturbations of the DE component, equivalent to a vanishing
of the combination (\ref{hatpH-1}):
\begin{equation}\label{holad}
\hat{p}_{H} =
\frac{\dot{p}_H}{\dot{\rho}_H}\hat{\rho}_{H}\quad
\Rightarrow\quad \hat{\omega} = - \frac{r \omega}{3}D_{H}\,.
\end{equation}
This assumption of an adiabatic DE component allows us to relate the otherwise undetermined perturbation $\hat{\omega}$ of the EoS parameter
to the DE energy perturbation $D_{H}$.
We emphasize that the total perturbation dynamics remains nonadiabatic due to the two-component nature of the medium.
The resulting coupled equations for $\delta^{c}$ and $S_{mH}$ in the $k$ space then are \cite{SRJW2} (the prime denotes a derivative with respect to $a$)
\begin{eqnarray}
\delta^{c\prime\prime} &+& \left[\frac{3}{2}-\frac{15}{2}\frac{p}{\rho}+ 3\frac{p^{\prime}}{\rho^{\prime}}\right]
\frac{\delta^{c\prime}}{a}\nonumber\\
&-& \left[\frac{3}{2} + 12\frac{p}{\rho} - \frac{9}{2}\frac{p^{2}}{\rho^{2}} - 9\frac{p^{\prime}}{\rho^{\prime}}
 - \frac{k^{2}}{a^{2}H^{2}}\frac{p}{\rho^{\prime}}\right]\frac{\delta^{c}}{a^{2}}\nonumber\\
 &&\qquad\qquad\qquad\qquad  = \frac{k^{2}}{a^{2}H^{2}}\frac{p^{\prime}}{\rho^{\prime}}\frac{\rho_{m}}{\rho}\frac{S_{mH}}{a^{2}}
\
  \label{prprdeltaS}
\end{eqnarray}
and
\begin{eqnarray}\label{prprS}
S^{\prime\prime}_{mH}
&+& \left[\frac{3}{2} - 3 \frac{r}{1+\omega}\frac{p^{\prime}}{
\rho^{\prime}} - \frac{3}{2}\frac{p}{\rho}\right]\frac{S^{\prime}_{mH}}{a}\nonumber\\
&+& \frac{r}{1+\omega}\frac{p}{\rho}\frac{k^{2}}{a^{2}H^{2}}\frac{S_{mH}}{a^{2}}
 = \frac{1+r}{1+\omega}\frac{p}{\rho}\frac{k^{2}}{a^{2}H^{2}}\frac{\delta^{c}}{a^{2}}\,,
\end{eqnarray}
respectively, where we have to exclude the case $\omega = -1$.

\subsection{Matter perturbations}

To obtain the matter-energy perturbations, we decompose
the total energy-density perturbation $\delta^{c}$ according to
\begin{equation}\label{}
\delta^{c} = \frac{\rho_{m}}{\rho}\delta_{m}^{c} + \frac{\rho_{H}}{\rho}\delta_{H}^{c} \,.
\end{equation}
Combination with $S_{mH} = \delta_{m} - \frac{\delta_{H}}{1+\omega}$ leads to
\begin{equation}\label{deltamc}
\delta_{m}^{c}= \frac{1}{1 + \frac{\omega}{1+r}}\left[\delta^{c} + \frac{1+\omega}{1+r}S_{mH}\right]\,,
\end{equation}
which describes the matter-energy perturbations as a combination of $\delta^{c}$ and $S_{mH}$. To obtain its dynamics one has to solve the coupled system of equations (\ref{prprdeltaS}) and (\ref{prprS}).

The matter density perturbation $\delta_{m}^{c}$ in relation (\ref{deltamc}) is defined with respect to the \textit{total} comoving
gauge. To obtain the matter density perturbation, comoving with the matter velocity,
$\delta_{m}^{c_{m}} = \delta_{m} + \frac{\dot{\rho}_{m}}{\rho_{m}}v_{m}$,
 we have to consider
\begin{equation}\label{}
\delta_{m}^{c} = \delta_{m} + \frac{\dot{\rho}_{m}}{\rho_{m}}v= \delta_{m}^{c_{m}}
+ \frac{\dot{\rho}_{m}}{\rho_{m}}\left(v-v_{m}\right)\,.
\end{equation}
Since \cite{SRJW2}
\begin{equation}\label{}
v-v_{m} = -\frac{\rho_{H} + p_{H}}{\rho+p}\frac{a^{2}}{k^{2}}\dot{S}_{mH}\,,
\end{equation}
the quantity of interest is
\begin{equation}\label{deltamcm}
\delta_{m}^{c_{m}} = \delta_{m}^{c}
- \frac{3}{1 + \frac{r}{1+\omega}}\frac{a^{2}H^{2}}{k^{2}}aS^{\prime}_{mH}\,.
\end{equation}
Obviously, $\delta_{m}^{c}$ and $\delta_{m}^{c_{m}}$ differ by the last term in relation (\ref{deltamcm}).
Because of the factor $\frac{a^{2}H^{2}}{k^{2}}$ (assuming $\omega \neq -1$) one expects that on scales smaller than the Hubble scale
the differences between $\delta_{m}^{c}$ and $\delta_{m}^{c_{m}}$ are small.

\subsection{A holographic model for the cosmic dynamics}

\begin{figure}[!t]
\begin{center}
\begin{minipage}[t]{0.50\linewidth}
\includegraphics[width=\linewidth]{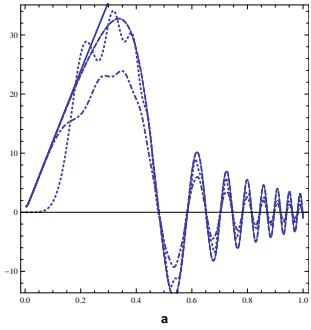}
\end{minipage} \hfill
\caption{Evolution of the perturbation quantities $S_{mH}$ (dotted line), $\delta^{c}$ (dash-dotted line),
$\delta^{c}_{m}$ (dashed line) and $\delta^{c_{_{m}}}_{m}$ (thin solid line)  for $\omega_{0} = -0.9$, $r_{0} = 0.4$ and $k= 0.1$. For comparison the $\Lambda$CDM result (thick solid line) is also included.}
\label{fig1}
\end{center}
\end{figure}

In Fig.~\ref{fig1} we show the behavior of the quantities  $S_{mH}$,  $\delta^{c}$, $\delta_{m}^{c}$ and $\delta_{m}^{c_{_{m}}}$ for $\omega_{0} = - 0.9$ with  $k=0.1$. While this figure confirms that differences between $\delta_{m}^{c}$ and $\delta_{m}^{c_{m}}$ are indeed small
on the chosen scale, there appear oscillations of all the perturbation quantities very close to the present time.
This reminds of a similar feature in (generalized) Chaplygin-gas models which apparently have jeopardized these models \cite{VDF}.
Still more serious is the existence of instabilities at future values $a > 1$  of the scale factor, related to a crossing of the phantom divide $\omega = -1$. Instabilities occur if the denominator $1 + \omega$ in (\ref{prprS}) vanishes, i.e., if $\omega$ approaches $-1$.
From Eq.~(\ref{wsol}) one finds the condition for $1 + \omega = 0$,
\begin{equation}\label{}
r_{0} = \omega_{0}\left(3 - \left(r_{0} - 3\omega_{0}\right)\right)a_{i}^{r_{0} - 3\omega_{0}}\ ,
\end{equation}
which determines the value $a_{i}$ of the scale factor at which the instability occurs. Solving for $a_{i}$
yields
\begin{equation}\label{}
a_{i}^{r_{0} - 3\omega_{0}}
= \frac{r_{0}}{\omega_{0}\left(3\left(1+\omega_{0}\right) - r_{0}\right)} \ .
\end{equation}
Now we assume $\omega_{0} = - 1 + \mu$ and consider
the cases $\omega_{0} > -1$ and $\omega_{0} < -1$ separately. For $\mu \neq \frac{r_{0}}{3}$ and $\mu \neq 1$
we have
\begin{equation}\label{}
a_{i}^{r_{0} - 3\omega_{0}} = \frac{r_{0}}{\left(r_{0} - 3\mu\right)\left(1 - \mu\right)}\ .
\end{equation}
For $\mu > 0$  we find $a_{i} > 1$, i.e., the instability sets in at a finite value of the scale factor
in the future.
For $\mu < 0$, i.e. for a phantom EoS, there appears an instability in the past
at $a_{i} < 1$. Since such kind of instability has not been observed, a present phantom EoS
is definitely excluded in the context of our model. The limit between the two regimes is just $\mu = 0$ where
we have $a_{i} = 1$, i.e., an instability at the present epoch.

The only case without instabilities at finite values of the scale factor is a fixed relation
$r_{0} - 3\omega_{0} = 3$ between the initially independent values of $r_{0}$ and $\omega_{0}$.
Since $r_{0} > 0$ necessarily, this implies $\omega_{0} > - 1$.
Consequently, the only physically acceptable case is
\begin{equation}\label{acceptable}
\omega_{0} = - 1 + \frac{r_{0}}{3}
\quad \Leftrightarrow\quad \Omega_{m0} = 3 \frac{1+\omega_{0}}{1+3\left(1+\omega_{0}\right)}\ .
\end{equation}
The parameters $\omega_{0}$ and $r_{0}$ are necessarily related to each other and cannot be chosen
independently.
In a sense, $r_{0}$ quantifies the deviation of $\omega_{0}$ from $\omega_{0} = - 1$. Under this condition we have $c^2 = \frac{1}{2}$. This is exactly the result found by Karwan and Thitapura in their study of instabilities through nonadiabatic perturbations in a system of matter and Ricci DE \cite{karwan}.

\begin{figure}[!t]
\begin{center}
\begin{minipage}[t]{0.50\linewidth}
\includegraphics[width=\linewidth]{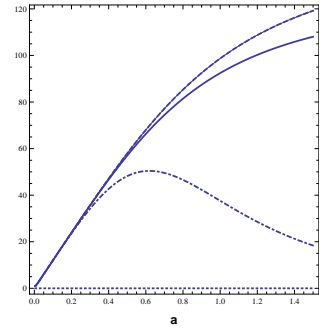}
\end{minipage} \hfill
\caption{Evolution of the perturbation quantities $S_{mH}$ (dotted line), $\delta^{c}$ (dash-dotted line),
$\delta^{c}_{m}$ (dashed line) and $\delta^{c_{_{m}}}_{m}$ (thin solid line) for $r_{0} = 0.4$ and $k= 0.1$ on the basis of (\ref{wsol+}) and (\ref{dotp0}).
The $\Lambda$CDM result is represented by the thick solid line. The relative density perturbations $S_{mH}$ are negligible during the entire evolution. The results for $\delta^{c}_{m}$ and $\delta^{c_{_{m}}}_{m}$ are almost identical.}
\label{fig2}
\end{center}
\end{figure}

The solutions for $\omega$ and $r$ then simplify to
\begin{equation}\label{wsol+}
\omega = \frac{\omega_{0}}{\left(1+\omega_{0}\right)a^{-3} - \omega_{0}}\ \mathrm{\ and}\ \
r = 3 \frac{r_{0}}{\left(3 - r_{0}\right)a^{3} +r_{0}}\,,
\end{equation}
respectively.
Combination of both solutions has the important consequence
\begin{equation}\label{dotp0}
\frac{r}{1 + \omega} =3\qquad \Rightarrow\qquad \frac{p^{\prime}}{\rho^{\prime}} = 0\,.
\end{equation}
This makes all the coupling terms (and some others) in the coupled system (\ref{prprdeltaS}) and (\ref{prprS}) vanish.
Also the pressure perturbations $\hat{p}^{c}$  vanish.
The square of the Hubble rate turns out to be
\begin{equation}\label{Hstab}
\frac{H^{2}}{H_{0}^{2}} = \Omega_{m0}a^{-3} + 1 + \frac{1}{3}\Omega_{m0}\left(a^{-3} - 4\right)\,.
\end{equation}
Notice that we have the same number of free parameters as in the $\Lambda$CDM model, but there is no
$\Lambda$CDM limit of (\ref{Hstab}).
The behavior of the perturbation quantities on the basis of (\ref{wsol+}) and (\ref{dotp0}) is visualized in Fig.~\ref{fig2}. This figure confirms that for the chosen configuration there are neither oscillations nor instabilities. From this point of view the model appears acceptable.

\begin{figure}[!t]
\begin{center}
\begin{minipage}[t]{0.50\linewidth}
\includegraphics[width=\linewidth]{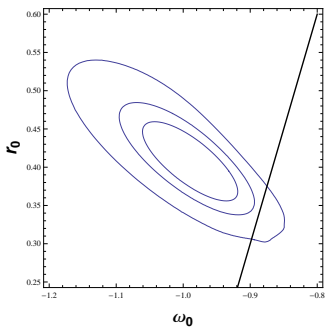}
\end{minipage} \hfill
\caption{Two-dimensional PDF for $\omega_0$ and $r_0$ resulting from a combination of the three tests.
The straight line represents the instability-avoiding configuration $r_{0} = 3\left(1 + \omega_0 \right)$.}
\label{fig3}
\end{center}
\end{figure}

We have tested this model by the differential age of old objects based on the
$H(z)$ dependence \cite{moresco,ratra} as well as by the data from SNIa \cite{union} and from BAO \cite{blake}. The details of the analysis are given in
Ref.~\onlinecite{SRJW2}. The results are presented in Fig.~\ref{fig3} which shows the
two-dimensional probability-distribution function (PDF) at $1\sigma$ ($68\%$ of confidence level), $2\sigma$ ($95\%$ of confidence level) and $3\sigma$ ($99\%$ of confidence level).
The estimation for $\omega_0$, based on a combination of the three tests at $2\sigma$, is $\omega_0 = - 0.987^{+0.083}_{-0.100}$, while for $r_0$ we find $r_0 = 0.406^{+0.073}_{-0.061}$.
The straight line represents the combination $r_{0} = 3\left(1 + \omega_0 \right)$ which is singled out by the stability analysis of the perturbation dynamics. The tension to the results for the background dynamics is obvious, an agreement is possible only at the $3\sigma$ level.

\section{\label{sec:summary}Summary}
Noninteracting Ricci-type DE is characterized by a necessarily time-dependent EoS parameter. This makes it an observationally testable alternative to the $\Lambda$CDM model.
There exists a a relationship between this EoS parameter and the matter content of the Universe.
 The ratio of the energy densities of DM and DE varies considerably less than for the
 $\Lambda$CDM model. Since the time of radiation decoupling it has changed by about one order of magnitude
 compared with roughly nine orders of magnitude for the  $\Lambda$CDM model. This amounts to a remarkable alleviation of the coincidence problem.
 Ricci-type DE behaves almost as dust at high redshift.
 Our statistical analysis, based on recent observational data from SNIa, BAO and $H(z)$, results in a preferred value of $c^{2}\approx 0.46$ for the Ricci-DE parameter which confirms earlier studies in the literature \cite{gao}.
  Within a gauge-invariant analysis we calculated the matter perturbations as a combination of the total energy perturbations of the cosmic medium and the relative perturbations of the components.
 The perturbation dynamics suffers from instabilities that exclude a present phantom-type EoS.
  It is only for a specific relation between the values $\Omega_{m0}$ of the present matter density and the present EoS parameter $\omega_{0}$ that the dynamics remains stable for any finite scale-factor value.
 This relation corresponds to a Ricci-DE parameter $c^{2}= 0.5$ \cite{karwan}.
  Holographic Ricci-type DE represents a theoretically appealing scenario which does not need additional parameters except $H_{0}$ and $\Omega_{m0}$. Despite of its  attractive features, the stable configuration is
  only marginally consistent with the observationally  preferred background values of $\Omega_{m0}$ and $\omega_{0}$.

\begin{acknowledgments}
 This work was supported by the ``Comisi\'{o}n
Nacional de Ciencias y Tecnolog\'{\i}a" (Chile) through the
FONDECYT Grants No. 1110230 and  No. 1130628 (R.H. and S.d.C).
J.C.F and W.Z acknowledge support by ``FONDECYT-Concurso incentivo a la
Cooperaci\'{o}n Internacional" No. 1130628 as well as by CNPq (Brazil) and FAPES (Brazil).
We thank Alonso Romero for carefully checking our calculations.
\end{acknowledgments}


\end{document}